\RequirePackage{ifpdf}
\ifpdf 
\documentclass[pdftex]{sigma}
\else
\documentclass{sigma}
\fi

\usepackage{accents}
\def \h#1{\widehat{#1}}
\def \t#1{\widetilde{#1}}
\def \b#1{\overline{#1}}
\def \c#1{\accentset{\circ}{#1}}
\def \c#1{\accentset{\circ}{#1}}
\def \th#1{\widehat{\widetilde{#1}}}
\def \hb#1{{\widehat{\overline{#1}}}}
\def \bh#1{{\widehat{\overline{#1}}}}
\def \bb#1{{\overline{\overline{#1}}}}
\def \tb#1{\widetilde{\overline{#1}}}

\def \dh#1{\underaccent{\hat}{#1}}
\def \db#1{\underaccent{\bar}{#1}}
\def \dt#1{\underaccent{\tilde}{#1}}

\def \dbb#1{\underaccent{\bar}{\underaccent{\bar}{#1}}}

\numberwithin{equation}{section}

\begin{document}


\renewcommand{\thefootnote}{$\star$}

\renewcommand{\PaperNumber}{046}

\FirstPageHeading

\ShortArticleName{Rational Solutions of the H3 and Q1 Models in the ABS Lattice List}

\ArticleName{Rational Solutions of the H3 and Q1 Models\\ in the ABS Lattice List\footnote{This
paper is a contribution to the Proceedings of the Conference ``Integrable Systems and Geomet\-ry'' (August 12--17, 2010, Pondicherry University, Puducherry, India). The full collection is available at \href{http://www.emis.de/journals/SIGMA/ISG2010.html}{http://www.emis.de/journals/SIGMA/ISG2010.html}}}

\Author{Ying SHI and Da-jun ZHANG}
\AuthorNameForHeading {Y.~Shi and D.J.~Zhang}
\Address{Department of Mathematics, Shanghai University, Shanghai 200444, P.R. China}
\Email{\href{mailto:shiying0707@shu.edu.cn}{shiying0707@shu.edu.cn}, \href{mailto:djzhang@staff.shu.edu.cn}{djzhang@staff.shu.edu.cn}}
\URLaddress{\url{http://www.science.shu.edu.cn/siziduiwu/zdj/index.htm}}

\ArticleDates{Received January 31, 2011, in f\/inal form May 04, 2011;  Published online May 09, 2011}

\Abstract{In the paper we present rational solutions for the H3 and Q1 models in the Adler--Bobenko--Suris lattice list.
These solutions are in Casoratian form and are generated by considering dif\/ference equation sets satisf\/ied by
the basic Casoratian column vector.}

\Classification{37K10}

\section{Introduction}

Recently, for many discrete systems which are integrable in the
sense of  multi-dimensional
consistency~\cite{Nijhoff-MDC,ABS-CMP-2003}, their explicit soliton
solutions were derived via constructive
approaches~\cite{Sol-Q4-2007,Sol-Q3-2008,NAJ,HZ-PartII,HZ-DBSQ-2010,Atkinson-CMP-2010,Nijhoff-INMS-2011}
such as the Cauchy Matrix approach and bilinear method. For a
continuous integrable system, once we have its multi-soliton
solutions, rational solutions usually can be found through limiting
procedures (cf.~\cite{Abl-Satsuma-1978,ZDJ-Wronskian}). In discrete
case, limiting procedures for getting  rational solutions are more
delicate and they often lead to a   trivialization of the  Cauchy
matrix. However, bilinear method has been successfully applied to
get limit solutions of some multi-dimensionally consistent lattices.
This was mentioned in~\cite{HZ-DBSQ-2010} for the lattice Boussinesq
equation and then discussed in detail in~\cite{ZH-H1} for the H1
equation in the Adler--Bobenko--Suris (ABS)'s list~\cite{ABS-CMP-2003}.
In fact, with the help of bilinear forms and Casoratians (see~\cite{HZ-PartII,HZ-DBSQ-2010,ZH-H1}), one can transfer a
lattice equation to a linear dif\/ference equation set satisf\/ied by a
$N$th-order Casoratian column vector. Usually there will be a
coef\/f\/icient matrix appearing in the dif\/ference equation set. When
the coef\/f\/icient matrix has $N$ distinct eigenvalues one gets
$N$-soliton solutions, and when it has~$N$ same eigenvalues one gets
limit solutions.

In this paper, we will see that
H3 and Q1 models in the ABS's list~\cite{ABS-CMP-2003} admit rational solutions.
In~\cite{HZ-PartII} Casoratian solutions for H3 and Q1 have been given with explicit Casoratian entries.
In this paper, we will f\/irst replace these explicit entries by the dif\/ference equation sets
satisf\/ied by them, by which it is possible to get more solutions beyond solitons.
We need to introduce a coef\/f\/icient matrix for each  dif\/ference equation set.
Then, by discussing eigenvalues of the coef\/f\/icient matrix we can get
several ``generalized'' solutions, and one of which is rational type.

The paper is organized as follows. Section~\ref{section2} provides basic notations for lattice variables, Casoratians and so forth.
In Section~\ref{section3} we derive rational solutions for H3 and in Section~\ref{section4} for~Q1.

\section{Preliminary}\label{section2}

Let us introduce some notations. Given a base point
$u_{n,m}=u$, we indicate the shifts in $n$, $m$ direction by up/down-tilde and up/down-hat, say
\begin{gather*}
  \widetilde{u}=u_{n+1,m}, \qquad \widehat{u}=u_{n,m+1} , \qquad \dt u=u_{n-1,m} , \qquad \dh u=u_{n,m-1}.
\end{gather*}
With these notations the lattice equations of our interest are
\begin{gather}
\mathrm{H3}\equiv
p  (u \t u+\h u \h{\t u})- q (u \h u+\t u \h{\t u})+\delta \big(p ^2-q^2\big) =0,
\label{H3}
\\
\mathrm{Q1}\equiv
p  (u-\h u) (\t u-\h{\t u})-
q (u-\t u) (\h u-\h{\t u})+\delta^2 p  q (p-q) =0,
\label{Q1}
\end{gather}
where $p$ and $q$ are independent parameters of $n$ and $m$
respectively, and $\delta$ is an arbitrary constants.
For generating soliton solutions, one needs to introduce an auxiliary direction $l$
which provides parameters of the solitons
by B\"{a}cklund transformation constructed from multi-dimensional consistency.
Shifts in $l$ direction are indicated by up/down-bar as follows, $\b u=u(l+1)$, $\db u=u(l-1)$.

Casoratian is a discrete version of Wronskian.
It is a determinant of a Casorati matrix composed of a basic column vector, i.e.,
\begin{gather*}
  |\psi(n,m,l_1),\psi(n,m,l_2),\dots,\psi(n,m,l_N)|=|l_1,l_2,\dots,l_N|,
\end{gather*}
where the basic column vector is
\begin{gather}
    \psi
    (n,m,l)=(\psi_1(n,m,l),\psi_2(n,m,l),\dots,\psi_N(n,m,l))^T.
\label{psi}
\end{gather}
Following the standard shorthand notation given in \cite{Freeman-Nimmo-KP}, some $N$th-order Casoratians
that we often use are indicated by
\begin{gather*}
|\h{N-1}|=|0,1,\dots,N-1|,\qquad  |\h{N-2},N|=|0,1,\dots,N-2,N|, \\ |-1,\t{N-1}|=|-1,1,2,\dots,N-1|.
\end{gather*}

The above Casoratians are def\/ined in terms of the shifts in $l$ direction. In fact, we can organize columns in terms of the shifts of~$n$ or~$m$.
By the operators $E^\nu$ $(\nu=1,2,3)$ we  denote the operations
\begin{gather*}
    E^1\psi=\widetilde{\psi}, \qquad E^2\psi=\widehat{\psi}, \qquad E^3\psi=\bar{\psi},
\end{gather*}
then we can def\/ine a Casoratian w.r.t.~$E^\nu$-shift,
\begin{gather*}
    |\widehat{N-1}|_{[\nu]}=\big|\psi,E^\nu\psi,(E^\nu)^2\psi,\dots,(E^\nu)^{N-1}\psi\big|, \qquad \nu=1,2,3.
\end{gather*}
We note that if the column vector $\psi$ satisf\/ies
\begin{gather*}
\b \psi=\t \psi+ \alpha \psi \qquad \mathrm{or} \qquad  \b \psi=\h \psi+ \beta \psi,
\end{gather*}
where $\alpha$ and $\beta$ are some constant, then we have
\begin{gather*}
    |\widehat{N-1}|_{[3]}=|\widehat{N-1}|_{[1]}\qquad \mathrm{or} \qquad |\widehat{N-1}|_{[3]}=|\widehat{N-1}|_{[2]}.
\end{gather*}

In addition to the above notations and properties,
we need the the following Laplace expansion identity for Casoratian verif\/ication.
\begin{lemma}[\cite{Freeman-Nimmo-KP}]
\label{lem-1}
Suppose that $\mathbf{B}$ is a  $N\times(N-2)$ matrix, and
\textbf{a}, \textbf{b}, \textbf{c}, \textbf{d} are $N$th-order column
vectors, then
\begin{gather*}
    |\mathbf{B},\mathbf{a},\mathbf{b}||\mathbf{B},\mathbf{c},\mathbf{d}|
    -|\mathbf{B},\mathbf{a},\mathbf{c}||\mathbf{B},\mathbf{b},\mathbf{d}|+|\mathbf{B},\mathbf{a},\mathbf{d}||\mathbf{B},\mathbf{b},\mathbf{c}|=0.
\end{gather*}
\end{lemma}

\section{Rational solutions for H3}\label{section3}

There are two dif\/ferent bilinear forms related to the H3 equation \eqref{H3}, which are~\cite{HZ-PartII}
\begin{subequations}
\label{eq:bil-H3}
\begin{gather}
\mathcal{B}_1  \equiv   2c f \t f +(a-c) \tb f \db f -(a+c)\b f
\db{\t f} =0,
\label{eq:bil-H3-1}\\
\mathcal{B}_2 \equiv   2c f \h f +(b-c) \hb f \db f -(b+c)\b f
\db{\h f} =0, \label{eq:bil-H3-2}
\end{gather}
\end{subequations}
and
\begin{gather}
  {\mathcal{B}}_1^{\prime}   \equiv    (b+c)\th f \b f
 +(a-c)f \th {\b f} -(a+b)\t f\bh f =0,\nonumber\\
 {\mathcal{B}}_2^{\prime}  \equiv    (c-b)\th f \db f
-(a+c)f \th {\db f} +(a+b)\t f\db{\h f} =0,\label{eq:bil-H3-2M1}\\
 {\mathcal{B}}_3^{\prime}  \equiv    (c-a)(b+c) \tb f
\db{\h f} +(a+c)(b-c) \hb f \db{\t f} +2c(a-b) f \th f =0,\nonumber
\end{gather}
via the same transformation
\begin{gather}\label{trans-H3}
  u= A \alpha^n \beta^m\frac{\b f}{f}+
  B \alpha^{-n}\beta^{-m}\frac{\db f}{f},\qquad AB=-\tfrac14r\delta,
\end{gather}
where the parametrization is
\begin{gather*}
    \frac{r^2 c^2}{c^2-a^2}=p^2,\qquad \frac{r^2 c^2}{c^2-b^2}=q^2,\qquad
    \alpha^2=-\frac{a-c}{a+c},\qquad \beta^2=-\frac{b-c}{b+c}.
\end{gather*}
Both bilinear forms can be solved by~\cite{HZ-PartII}
\begin{gather}
f(\psi)=|\h{N-1}|_{[3]}
\label{cas-f}
\end{gather}
with the Casoratian column vector $\psi$ (see the structure \eqref{psi}) composed of
\begin{gather}\label{1.1}
  \psi_i(n,m,l)=\rho_i^+(a+k_i)^n(b+k_i)^m(c+k_i)^l+\rho_i^-(a-k_i)^n(b-k_i)^m(c-k_i)^l,
\end{gather}
where $\rho_i^{\pm}$ and $k_i$ are parameters.

Now we consider a generalization of the above $\psi$.
We will discuss the linear dif\/ference equation set satisf\/ied by $\psi$.
First, there should exist symmetric relationship between $(n,a)$ pair and $(m,b)$ pair.
This can be seen from the symmetric position of $n$ and $m$ appearing in the H3 equation.
In addition, from the bilinear forms \eqref{eq:bil-H3} and \eqref{eq:bil-H3-2M1}
this symmetric relationship can be extended to include $(l,c)$ pair.
Since $\psi$ plays a role of the basic Casoratian column vector, it will consequently keep the $n$-$m$-$l$ symmetric property.

In the generalization procedure, Toeplitz matrices play important
roles. A $N$th-order lower triangular Toeplitz matrix (LTTM)
$\mathcal{F}$ is a matrix of the form (see \cite{ZDJ-Wronskian} for
more properties)
\begin{gather}\label{LTTM-F}
    \mathcal{F}=\left(
                       \begin{array}{cccccc}
                        f_0&0 &0 &\cdots  & 0 & 0 \\
                        f_1 & f_0  & 0 & \cdots &0 & 0 \\
                       f_2 & f_1 & f_0 &\cdots & 0 & 0 \\
                       \cdots &\cdots  & \cdots & \cdots & \cdots & \cdots  \\
                       f_{N-1} & f_{N-2}  & f_{N-3} &\cdots &f_1 & f_0  \\
                       \end{array}
                     \right),\qquad f_j\in \mathbb{C}.
\end{gather}
$\mathcal{F}(k)$ is called a $N$th-order LTTM generated from the function $F(k)$, if in \eqref{LTTM-F}
\[
f_j=\frac{1}{j!}\partial^{j}_k F(k),
\]
where the `seed' function $F(k)$ is arbitrarily dif\/ferentiable w.r.t.~$k$.
\begin{lemma}
\label{lem-2}
Let non-zero functions $F(k,l)$ and $G(k)$ be sufficiently smooth w.r.t.~$k$ and satisfy
\[\partial_l F(k,l)=F(k,l)G(k).\]
Here $l$ is an auxiliary argument of $F(k,l)$.
If $\mathcal{F}(k,l)$ and $\mathcal{G}(l)$ are respectively two $N$th-order LTTMs generated from
$F(k,l)$ and $G(k)$, then we have
\begin{gather}
\partial_l \mathcal{F}(k,l)= \mathcal{F}(k,l) \mathcal{G}(k).
\label{FG}
\end{gather}
\end{lemma}

\begin{proof}
\begin{gather*}
\partial_l f_j  = \frac{1}{j!}\partial^{j}_k \partial_l F(k,l)
                = \frac{1}{j!}\partial^{j}_k (F(k,l)G(k))
                = \frac{1}{j!}\sum^{j}_{s=0}C^s_{j}(\partial^{j-s}_k F(k,l))(\partial^{s}_k G(k))\\
\phantom{\partial_l f_j}{}
                = \sum^{j}_{s=0}\left(\frac{1}{(j-s)!} \partial^{j-s}_k F(k,l)\right)\left(\frac{1}{s!}\partial^{s}_k G(k)\right)
                = \sum^{j}_{s=0}f_{j-s}\cdot g_s.
\end{gather*}
This yields the relation \eqref{FG}.
\end{proof}
\begin{corollary}
\label{coro-lem2}
Under the condition of Lemma~{\rm \ref{lem-2}}, the product of LTTMs $\mathcal{F}(k,l) \mathcal{F}^{-1}(k,l+1)$   is independent of $l$.
\end{corollary}

\begin{proof}
From  \eqref{FG} we have
\[
\mathcal{F}^{-1}(k,l)\cdot \partial_l  \mathcal{F}(k,l)=\mathcal{G}(k),
\]
which means $\mathcal{F}^{-1}(k,l)\cdot \partial_l  \mathcal{F}(k,l)$ is independent of $l$,
i.e.,
\[
\mathcal{F}^{-1}(k,l)\cdot \partial_l  \mathcal{F}(k,l)=  \mathcal{F}^{-1}(k,l+1)\cdot \partial_l \mathcal{F}(k,l+1).
\]
Noting that
\begin{gather*}
  \partial_l  (\mathcal{F}(k,l) \mathcal{F}^{-1}(k,l+1))\\
\qquad{} =  (\partial_l  \mathcal{F}(k,l))  \mathcal{F}^{-1}(k,l+1)-\mathcal{F}(k,l) \mathcal{F}^{-1}(k,l+1)
 (\partial_l \mathcal{F}(k,l+1)) \mathcal{F}^{-1}(k,l+1)
\end{gather*}
and then replacing $\mathcal{F}^{-1}(k,l+1)\cdot \partial_l \mathcal{F}(k,l+1)$ by
$\mathcal{F}^{-1}(k,l)\cdot \partial_l  \mathcal{F}(k,l)$ yield
\[ \partial_l  (\mathcal{F}(l) \mathcal{F}^{-1}(l+1))=0.\]
Thus we complete the proof.
\end{proof}

Now let us come to the main results of H3.
\begin{theorem}
\label{t-1}
For {\rm H3} the Casoratian $f(\psi)$ defined in \eqref{cas-f} solves the bilinear forms \eqref{eq:bil-H3} and~\eqref{eq:bil-H3-2M1}
if $\psi$ satisfies
\begin{subequations}  \label{eq:cas-for-H3}
\begin{gather}
\b \psi=\t \psi +(c-a)\psi,
\label{cd-h3-1}
\end{gather}
and there is a $N$th-order auxiliary vector $\sigma(n,m,l)$ such
that
\begin{gather}
\psi=A_{[l]}\, \sigma,
\label{psi-sigma}
\end{gather}
and
\begin{gather}
\db\sigma=-\h\sigma+(b+c)\sigma,
\label{sigma-shift}
\end{gather}
\end{subequations}
where $A_{[l]}$ is a $N$th-order transform matrix, $A_{[l]}
A^{-1}_{[l+1]}$ is independent of $l$, and the subscript $[l]$
specially means that $A_{[l]}$ only depends on $l$ but is
independent of $(n,m)$. Moreover, in the light of $n$-$m$-$l$
symmetric property and \eqref{psi-sigma}, the following relation
\[\psi=A_{[n]}  \omega, \qquad \psi=A_{[m]}  \phi\]
automatically holds,
where  $\omega(n,m,l)$ and $\phi(n,m,l)$ are auxiliary vectors.
\end{theorem}

\begin{proof}
The proof is similar to the one with concrete $\psi_i$ \eqref{1.1} in~\cite{HZ-PartII}, but here we need to carefully examine
the role played by the transform matrix~$A_{[l]}$.
We only take~\eqref{eq:bil-H3-1} as an example.
By a~down-tilde-shift it goes to
\begin{gather}
2c \dt{f} f +(a-c) \b f \undertilde{\db f} -(a+c)\dt{\b f} \db f =0.
\label{eq:bil-H3-1d}
\end{gather}
The shift relation \eqref{cd-h3-1} and its symmetric forms suggest
\[ f(\psi)=|\h{N-1}|_{[i]}=|\h{N-1}|_{[j]},\qquad i,j=1,2,3.\]
So for \eqref{eq:bil-H3-1d} we consider $ f(\psi)=|\h{N-1}|_{[2]}$ for convenience.
Same as the derivation in~\cite{HZ-PartII}, we have
\begin{gather*}
  -(a-b)^{N-2}\dt f(\psi)  =  |\h{N-2},\dt \psi(N-2)|_{[2]},\qquad
  -(c-b)^{N-2}\db f(\psi)  =  |\h{N-2},\db \psi(N-2)|_{[2]},
\end{gather*}
and
\begin{gather*}
   (a-c)(c-b)^{N-2}(a-b)^{N-2} \undertilde{\db f}(\psi)=|\h{N-3},\db\psi(N-2),\dt\psi(N-2)|_{[2]}.
\end{gather*}
Besides, with the help of the shift relation \eqref{sigma-shift} for $\sigma$, we have
\begin{gather}
(c+b)^{N-2}\b f(\sigma)=|\sigma(0),\sigma(1),\dots,\sigma(N-2), \b\sigma(N-2)|_{[2]}.
\label{x1}
\end{gather}
Then, noting that $f(\psi)=|A_{[l]}|f(\sigma)$ in terms of $m$-shift construction of Casoratians
and def\/ining
\begin{gather*} 
 \c E^3\psi  = A_{[l]} A^{-1}_{[l+1]} E^3 \psi,
\end{gather*}
we f\/ind from \eqref{x1} that
\begin{gather} (c+b)^{N-2}\b f(\psi)=|A_{[l+1]}||A^{-1}_{[l]}||\h{N-2},\c E^3\psi(N-2)|_{[2]},
\label{f-bar}
\end{gather}
and further
\[ (a+c)(c+b)^{N-2}(a-b)^{N-2} \b{\undertilde
   f}(\psi)=|A_{[l+1]}||A^{-1}_{[l]}||\h{N-3},\dt\psi(N-2),\c E^3\psi(N-2)|_{[2]}.\]
Imposing a down-bar shift on \eqref{f-bar} and using the relation \eqref{cd-h3-1} yield
\[ 2c(c-b)^{N-2}(c+b)^{N-2}f(\psi)= |A_{[l+1]}||A^{-1}_{[l]}||\h{N-3},\db\psi(N-2),\c E^3\psi(N-2)|_{[2]},\]
where we have made use of the assumption of $A_{[l]} A^{-1}_{[l+1]}$
being independent of $l$. Now collecting these obtained formulae,
substituting them into the l.h.s.\ of \eqref{eq:bil-H3-1d} and using
Lemma~\ref{lem-1}, we can prove~\eqref{eq:bil-H3-1d}. The proof of
other formulae in the bilinear forms~\eqref{eq:bil-H3} and~\eqref{eq:bil-H3-2M1} are similar and we do not give further
details.
\end{proof}

Now let us see what is new the theorem brings. One choice satisfying
Theorem~\ref{t-1} is to take~$\psi$ which is composed of~\eqref{1.1},  $\sigma$ composed of
\begin{gather*}
    \sigma_i(n,m,l)=\rho_i^+(a+k_i)^n(b+k_i)^{m}(c-k_i)^{-l}+\rho_i^-(a-k_i)^n(b-k_i)^{m}(c+k_i)^{-l},
\end{gather*}
and
\begin{gather}\label{eq:T}
  A_{[l]}=\mbox{Diag}(A(k_1,l),A(k_2,l),\dots,
    A(k_N,l)),\qquad A(k_j,l)=\big(c^2-k_j^2\big)^l.
\end{gather}
This is nothing but the result given in~\cite{HZ-PartII} and it generates $N$-soliton solutions for H3.

Now let $A_{[l]}$ be a $N$th-order LTTM in the form of
\eqref{LTTM-F}, which is generated from $A(k_1,l)=(c^2-k_1^2)^l$,
i.e.,
\begin{gather*}
a_j=\frac{1}{j!}\partial_{k_1}^{j}\big(c^2-k_1^2\big)^l.
\end{gather*}
Then, the desirable Casoratian column vector $\psi$ can be taken as
\begin{gather*}
    \psi(n,m,l)=\mathcal{A}_+\psi^+(n,m,l)+\mathcal{A}_-\psi^-(n,m,l),
\end{gather*}
where
\begin{gather*}
\psi^{\pm}(n,m,l) = \big(\psi^\pm_1(n,m,l),\psi^\pm_2(n,m,l),\dots,\psi^\pm_{N}(n,m,l)\big)^T,\\
\psi^\pm_i(n,m,l) = \frac{1}{(i-1)!}\partial_{k_1}^{i-1}\big[\rho_1^{\pm}(a\pm
k_1)^{n}(b\pm k_1)^{m}(c\pm k_1)^{l}\big],
\end{gather*}
and $\mathcal{A}_{\pm}$ are two arbitrary non-singular LTTMs. The corresponding auxiliary vector $\sigma$ is given by
\begin{gather*}
    \sigma=\mathcal{A}_+\sigma^+(n,m,l)+\mathcal{A}_-\sigma^-(n,m,l),
\end{gather*}
where
\begin{gather*}
\sigma^{\pm}(n,m,l) = \big(\sigma^\pm_1(n,m,l),\sigma^\pm_2(n,m,l),\dots,\sigma^\pm_{N}(n,m,l)\big)^T,\\
\sigma^\pm_i(n,m,l) = \frac{1}{(i-1)!}\partial_{k_1}^{i-1}\big[\rho_1^{\pm}(a\pm
k_1)^{n}(b\pm k_1)^{m}(c\mp k_1)^{l}\big].
\end{gather*}
We note that  in the light of Corollary~\ref{coro-lem2},
$A_{[l]} A^{-1}_{[l+1]}$ is independent of $l$.
Besides, to avoid the generation of zero by high order derivatives
we may let $l\geq N$ or $l\not\in \mathbb{Z}^+$.
(This can  be done, for example, by taking
$(c\pm k_1)^{l_0}\rho_1^{\pm}$ in place of the original $\rho_1^{\pm}$
where $l_0$ is either a positive integer greater than $N-1$ or not a positive integer.
As an example, see \eqref{psi3} where we have taken $l_0=2$.)

As the simplest nontrivial case we take $N=2$ and thus the Casoratian $f$ is in the form of
\begin{subequations}
\begin{gather}
    f=|\psi(n,m,0),\psi(n,m,1)|_{[3]}
\label{H3-reslut}
\end{gather}
where
\begin{gather}
  \psi(n,m,l)=(\psi_1(n,m,l),\partial_{k_1}\psi_1(n,m,l))^T,\label{psi1} \\
  \psi_1(n,m,l)=\psi_1^+(n,m,l)+\psi_1^-(n,m,l),\label{psi2} \\
  \psi_1^{\pm}(n,m,l)=\rho_1^{\pm}(a\pm
k_1)^{n}(b\pm k_1)^{m}(c\pm k_1)^{2+l}
\label{psi3},
\end{gather}
\end{subequations}
and we have taken $\mathcal{A}_{\pm}=I$.
In this case, the coef\/f\/icient (or transform) matrix $A_{[l]}$ is a 2nd-order LTTM  generated from $A(k_1,l)=(c^2-k_1^2)^{2+l}$, i.e.,
\begin{gather*}
A_{[l]}=\left|\begin{array}{cc}
(c^2-k_1^2)^{2+l}&0\vspace{1mm}\\
\partial_{k_1}(c^2-k_1^2)^{2+l}& (c^2-k_1^2)^{2+l}
\end{array}\right|.
\end{gather*}
Then, substituting \eqref{H3-reslut} into the transformation~\eqref{trans-H3} yields a solution to~H3,
which is dif\/ferent from 2-soliton solution,
\begin{subequations}\label{H3-s}
\begin{gather}
  u_{n,m} = A \alpha^n \beta^m \frac{(c-k_1)^2-[D_1+4k_1 c(c^2-k_1^2)^{-1}
  ](c^2-k_1^2)Q_{n,m}-(c+k_1)^2Q_{n,m}^2}
  {1 - D_1Q_{n,m} - Q_{n,m}^2}\nonumber\\
{}+ B \alpha^{-n}\beta^{-m} \frac{(c-k_1)^{-2}\!-[D_1\!-4k_1c(c^2\!-k_1^2)^{-1}
  ](c^2\!-k_1^2)^{-1}Q_{n,m}\!-(c+k_1)^{-2}Q_{n,m}^{2}}
  {1 - D_1Q_{n,m} - Q_{n,m}^2},\!\!\!
\end{gather}
where
\begin{gather}
Q_{n,m} =\left(\frac{c+k_1}{c-k_1}\right)^2\left(\frac{a+k_1}{a-k_1}\right)^n\left(\frac{b+k_1}{b-k_1}\right)^m\rho_{0,0},\label{Qnm}\\
  D_1 =\frac{4 k_1[a
(b^2-k_1^2)(c^2-k_1^2)n+b(a^2-k_1^2)(c^2-k_1^2)m+2c(a^2-k_1^2)(b^2-k_1^2)]}{(a^2-k_1^2)(b^2-k_1^2)(c^2-k_1^2)},\label{D1}\\
 A B  =-\tfrac14r\delta,
\end{gather}
\end{subequations}
and the constants $\rho^{\pm}_1$ have been absorbed into the
parameter $\rho_{0,0}$.
Comparing with
the previous results in \cite{NAJ,HZ-PartII}, the function $D_1$  which
depends on~$n$, $m$ is new, and this is the generalization.

Recalling the transform(or coef\/f\/icient) matrix $A_{[l]}$ def\/ined in \eqref{eq:T},
which is a diagonal and has $N$ distinct eigenvalues. It leads to
$N$-soliton solutions. When $A_{[l]}$ is the $N$th-order LTTM
generated from $A(k_1,l)=(c^2-k_1^2)^l$, which is a matrix with
$N$ same eigenvalues $A(k_1,l)$, we get a kind of solutions which is
closely related to some limiting procedures (see
\cite{ZDJ-Wronskian,ZH-H1}) and dif\/ferent from solitons. Such
solutions we call limit solutions. Usually taking $k_1=0$ in such
limit solutions one may get rational solutions. In practice, we may
take $\psi_1$  to be an even function of~$k_1$, for example,
\begin{gather}\label{R1}
\psi_1(n,m,l)=(a+k_1)^{n}(b + k_1)^{m}(c +
k_1)^{l_0+l}+(a-k_1)^{n}(b - k_1)^{m}(c -k_1)^{l_0+l},
\end{gather}
where $l_0$ is some constant as we mentioned before.
Correspondingly, a nontrivial solution of \eqref{cd-h3-1} can be taken as
\begin{gather}\label{R}
\psi(n,m,l)=\left(\psi_1(n,m,l),\frac{1}{2!}\partial_{k_1}^{2}\psi_1(n,m,l),\dots,\frac{1}{(2N-2)!}\partial_{k_1}^{2N-2}\psi_1(n,m,l)\right)^T,
\end{gather}
and the transform matrix $A_{[l]}$ is a LTTM in the form of \eqref{LTTM-F} with
\begin{gather*}
a_j=\frac{1}{(2j)!}\partial_{k_1}^{2j}\big(c^2-k_1^2\big)^{l_0+l},\qquad j=0,1,\dots, N-1.
\end{gather*}
Then, the Casoratian for rational solutions is
\begin{gather*}
f=|\psi(n,m,0),\psi(n,m,1),\dots,\psi(n,m,N-1)|_{k_1=0}.
\end{gather*}
The following is a rational solution (with $N=2$ and $l_0=2$),
\begin{gather}
  u= A \alpha^n \beta^m c^2\left(1+\frac{ab}{c(b n + a m) + 2ab}\right)+
  B \alpha^{-n}\beta^{-m}c^{-2}\left(1-\frac{ab}{c(b  n + a m) + 2ab}\right),
\label{H3-r}
\end{gather}
where $AB=-\tfrac14r\delta$.

\section{Rational solutions for Q1}\label{section4}

\subsection{For bilinearization-I}

There are two types of bilinear forms for Q1 equation~\eqref{Q1}. One is~\cite{HZ-PartII}
\begin{subequations}\label{Q1-bil-I}
\begin{gather}
\mathcal Q_1 \equiv  {\h{\tb f}}  f  (b-\delta )+
\h{\t f} \b{f}  (a+\delta )-{\tb{f}} \h{f} (a+b)=0,\label{Q1-lin-bileqs-1}\\
\mathcal Q_2 \equiv   {\h{\tb f}} f (a-b)
+{\tb{f}} \h{f} (b+\delta )-\t{f}  {\hb{f}} (a+\delta )=0,\label{Q1-lin-bileqs-2}\\
\mathcal Q_3 \equiv  - {\tb{f}} \h{f}+ {\tb{f}} \h{\mathfrak{g}} (-a+\delta )
+\t{f}  {\hb{f}}+ {\hb{f}} \t{\mathfrak{g}} (b-\delta)
+\b{f} \h{\t{\mathfrak{g}}} (a-b)=0,\label{Q1-lin-bileqs-3}\\
\mathcal Q_4 =  {\h{\tb f}}\mathfrak{g} (a-b)+ {\tb{f }} \h{\mathfrak{g}} (a+b) - {\hb{f
}}\t{\mathfrak{g}} (a+b)+\b{f} \h{\t{\mathfrak{g}}} (-a+b)=0,\label{Q1-lin-bileqs-4}
\end{gather}
\end{subequations}
where the transformation is
\begin{gather}
u  =\alpha n+\beta m + \gamma -\big(c^2/r-\delta^2 r\big) \frac{\mathfrak{g}}{f}
\label{trans-Q1-lin}
\end{gather}
with parametrization
\begin{gather*}
\alpha=pa,\qquad \beta=qb,\qquad \frac{c^2/r-\delta^2
r}{a^2-\delta^2}=p,\qquad \frac{c^2/r-\delta^2 r}{b^2-\delta^2}=q.
\end{gather*}

By examining the Casoratian verif\/ication we f\/ind solutions of Q1 admit  the following gene\-ra\-li\-zation.

\begin{theorem}
\label{t-2}
The bilinear {\rm Q1} \eqref{Q1-bil-I} can be solved by
\begin{gather*}
  f(\psi)=|\h{N-1}|_{[3]},\qquad  \mathfrak{g}(\psi)=|-1,\t{N-1}|_{[3]},
\end{gather*}
if $\psi$ satisfies the shift relation
\begin{gather*}
\b \psi=\t \psi +(\delta-a)\psi
\end{gather*}
as well as $n$-$m$-$l$-symmetric property,
and there are $N$th-order auxiliary vector $\sigma(n,m,l)$ and $\phi(n,m,l)$ such that
\begin{gather*}
\psi=A_{[l]}  \sigma, \qquad \b\sigma=-\dh \sigma+(b+\delta)\sigma,
\end{gather*}
and
\begin{gather*}
\psi=A_{[n]}A_{[m]}  \phi, \qquad \b\phi=-\dt\phi+(a+\delta)\phi,\qquad \b\phi=-\dh\phi+(b+\delta)\phi,
\end{gather*}
where $A_{[l]}$ is defined as in Theorem~{\rm \ref{t-1}}, $A_{[l]} A^{-1}_{[l+1]}$ is independent of~$l$,
$A_{[n]}$ and $A_{[m]}$ posses pro\-per\-ties similar to $A_{[l]}$.
\end{theorem}

The proof is similar to the one in the previous section for H3 and in~\cite{HZ-PartII} for~Q1.
We skip it.

$N$-soliton solutions\cite{HZ-PartII} can be derived from those Casoratians by taking the transform matrixes as
\begin{gather*}
    A_{[\mu]}=\mbox{Diag}(A(k_1, \mu),A(k_2, \mu),\dots,
    A(k_N,\mu)),\qquad A(k_j,\mu)=\big(x_\mu^2-k_j^2\big)^\mu,
\end{gather*}
where
\[\mu=n,m,l,\qquad x_1=a, \qquad x_2=b, \qquad x_3=\delta,\]
and the basic column vector $\psi$ composed of
\begin{gather}
\psi_i(n,m,l)= \rho_{i}^{+}(a +k_i)^n(b +k_i )^m(\delta+ k_i)^l+
  \rho_{i}^{-}(a -k_i)^n(b -k_i )^m(\delta-k_i)^l.
  \label{psi-Q1-lin}
\end{gather}

As a generalization, $A_{[\mu]}$ can be  a LTTM generated from
$A(k_1,\mu)$. In this case, $\psi$ can be taken as
\begin{subequations}
\label{psi-Q1-lin-2}
\begin{gather}
    \psi(n,m,l)=\mathcal{A}_+\psi^+(n,m,l)+\mathcal{A}_-\psi^-(n,m,l),
\end{gather}
with
\begin{gather}
\psi^{\pm}(n,m,l) = \big(\psi^\pm_1(n,m,l),\psi^\pm_2(n,m,l),\dots,\psi^\pm_{N}(n,m,l)\big)^T,\\
\psi^\pm_i(n,m,l) = \frac{1}{(i-1)!}\partial_{k_1}^{i-1}\big[\rho_1^{\pm}(a\pm
k_1)^{n}(b\pm k_1)^{m}(\delta \pm k_1)^{l_0+l}\big],
\end{gather}
\end{subequations}
and arbitrary $N$th-order LTTMs $\mathcal{A}_{\pm}$.
Here we add the parameter $l_0$ which, as before, plays a~role of avoiding triviality of some high order derivatives.
Rational solutions may come out by  taking $k_1=0$ in the above $ \psi(n,m,l)$.

As examples we give two solutions which are not soliton solutions.
The f\/irst one is
\begin{gather*}
     u_{n,m}=\alpha n+\beta
    m+\gamma+2\big(c^2/r-\delta^2r\big)\frac{(\delta+k_1)-[\delta E_1-2k_1(\delta^4\!-k_1^4)]Q_{n,m}-(\delta-k_1)Q_{n,m}^2}
    {(\delta^2-k_1^2)(1-E_1Q_{n,m}-Q_{n,m}^2)},
\end{gather*}
where $Q_{n,m}$ and $E_1$ are given in \eqref{Qnm} and \eqref{D1}
with  $\delta$ in place of $c$. This solution is derived from the
transformation \eqref{trans-Q1-lin} in which we use the 2nd-order Casoratians
\[f=|\psi(n,m,0),\psi(n,m,1)|,\qquad g=|\psi(n,m,-1),\psi(n,m,1)|\]
with column vector~\eqref{psi-Q1-lin-2} where $\mathcal{A}_{\pm}=I$ and $l_0=2$.
Another example is a rational solution,
\begin{gather*}
    u_{n,m}=\alpha n+\beta m+\gamma-\delta^{-1}(c^2/r-\delta^2r)\left(2-\frac{ab}{\delta(b  n
    +a  m)
    +2ab}\right).
\end{gather*}
This is derived from the  2nd-order Casoratians
\[f=|\psi(n,m,0),\psi(n,m,1)|_{k_1=0},
\qquad g=|\psi(n,m,-1),\psi(n,m,1)|_{k_1=0},\]
where the  basic vector column $\psi(n,m,l)$ is~\eqref{R}
with even generator~\eqref{R1}, $N=2$ and $l_0=2$,
which is the same as  that we used to generate rational solution~\eqref{H3-r} for~H3.

\subsection{For bilinearization-II}

The second bilinearlization for Q1 is derived through
the transformation \cite{HZ-PartII}
\begin{gather*}
  u   = A \alpha^n \beta^m\frac{\bb f}{f}+
  B \alpha^{-n}\beta^{-m}\frac{\dbb f}{f},\qquad AB=\delta^2r^2/16,
\end{gather*}
with the parametrization is
\begin{gather*}
-\tfrac14r(1-\alpha)^2/\alpha=p, \qquad -\tfrac14r(1-\beta)^2/\beta=q.
\end{gather*}
The bilinear form is exactly the same as~\eqref{eq:bil-H3}, which is a bilinear H3.
In this case~Q1 can share Theorem~\ref{t-1} and the Casoratian column vector $\psi$ with H3, as given in Section~\ref{section3}.
By the same Casoratians $f$ that we used to generate solutions~\eqref{H3-s} and~\eqref{H3-r} in Section~\ref{section3},
the corresponding solutions of Q1 are
\begin{gather*}
     u_{n,m}=A
    \alpha^n\beta^m\frac{(c-k_1)^{4}-[(c^2-k_1^2)D_1+8k_1c](c^2-k_1^2)Q_{n,m}-(c+k_1)^{4}Q_{n,m}^2}{1-D_1Q_{n,m}-Q_{n,m}^2}\\
\phantom{u_{n,m}=}{} +B
    \alpha^{-n}\beta^{-m}\frac{(c-k_1)^{-4}-[(c^2-k_1^2)D_1-8k_1c](c^2-k_1^2)^{-3}Q_{n,m}-(c+k_1)^{-4}Q_{n,m}^2}{1-D_1Q_{n,m}-Q_{n,m}^2},
\end{gather*}
and
\begin{gather}
  u_{n,m}= A \alpha^n \beta^m c^4\left(1+\frac{2ab}{c(b\,n + a\,m) + 2ab}\right)\nonumber\\
  \phantom{u_{n,m}=}{} +
  B \alpha^{-n}\beta^{-m}c^{-4}\left(1-\frac{2ab}{c(b\,n + a\,m) + 2ab}\right).\label{Q1-r2}
\end{gather}
where $AB=\delta^2r^2/16$, $Q_{n,m}$ and $D_1$ are respectively def\/ined in~\eqref{Qnm} and~\eqref{D1}. \eqref{Q1-r2}~gives a rational solution of~Q1.

\section{Conclusions}\label{section5}

We have shown that H3 and Q1 models in the ABS's list admit more
solutions  than solitons. The coef\/f\/icient(or transform) matrix
$A_{[\mu]}$ in the Casoratian conditions, i.e., those dif\/ference
equation sets for the Casoratian column vector $\psi$, plays an
important role. When $A_{[\mu]}$ is a diagonal matrix with $N$
distinct eigenvalues, we get $N$-soliton solutions. $A_{[\mu]}$ can
also be a LTTM with~$N$ same eigenvalues. In this case, one gets
solutions dif\/ferent from solitons. Particularly, $k_1=0$ generates
rational solutions. $A_{[\mu]}$ can also be a combination of
diagonal blocks and LTTM blocks. We note that in continuous cases
these LTTM-type solutions (related to LTTMs) can be considered as
limit solutions of solitons (cf.~\cite{ZDJ-Wronskian}), and this is
also true for discrete cases. H1~admits LTTM-type solutions~\cite{ZH-H1} but it
does not have rational solutions. The reason is that an invertible
$A_{[\mu]}$ is needed in the Casoratian proof for~H1, but this will be broken
if taking $k_1=0$. H2~has the same situation as~H1. In fact, for H1
and H2, the basic Casoratian entry is
\[\psi_i(n,m,l)= \rho_{i}^{+}(a +k_i)^n(b +k_i )^m  k_i^l+
  \rho_{i}^{-}(a -k_i)^n(b -k_i )^m( -k_i)^l,\]
i.e., $c=0$ in \eqref{1.1} for~H3, or $\delta=0$ in
\eqref{psi-Q1-lin} for~Q1. However, for~H3 and~Q1, it is just the
existence of $c$ and $\delta$ to guarantee the non-triviality of $A_{[\mu]}$
when $k_1$ goes to zero. To keep $A_{[\mu]}$ invertible is also the
criterion to examine rational solution reduction ($\delta=0$) for~H3
and~Q1. Obviously, the rational solutions derived from the f\/irst
bilinear form of~Q1 do not admit reduction of $\delta=0$, but others
do. For~H1 and~H2, trying to introduce an auxiliary parameter (for
example, $c$ for~H3) in their Casoratian entry so that the transform
matrix~$A_{[\mu]}$ is still invertible as~$k_1\to 0$ might be a~possible way to get their rational solutions in Caosratian form.
This is left for further discussion.

\subsection*{Acknowledgements}

The authors are very grateful to the referees for their invaluable comments.
This project is supported by the NSF of China
(11071157) and Shanghai Leading Academic Discipline
Project (No.~J50101).

\pdfbookmark[1]{References}{ref}
\LastPageEnding

\end{document}